\documentclass[twocolumn,showpacs]{revtex4}

\usepackage{graphicx}%
\usepackage{dcolumn}
\usepackage{amsmath}
\usepackage{here}

\begin{document}

\title{Nonlinear dual-core photonic crystal fiber couplers}

\author{Jos\'e R. Salgueiro$^{\dagger}$ and Yuri S. Kivshar}

\affiliation{Nonlinear Physics Centre and Center for Ultra-high
bandwidth Devices for Optical Systems (CUDOS), Research School of
Physical Sciences and Engineering, The Australian National
University, Canberra ACT 0200, Australia}

\begin{abstract}
We study nonlinear modes of dual-core photonic crystal fiber
couplers made of a material with the focusing Kerr nonlinearity.
We find numerically the profiles of symmetric, antisymmetric, and
asymmetric nonlinear modes, and analyze all-optical switching
based on instability of the symmetric mode. We also describe {\em
elliptic spatial solitons} controlled by the waveguide boundaries.
\end{abstract}


\maketitle

Photonic crystal fibers (PCF) have attracted a lot of attention
due to their intriguing properties, potential applications, and
the development of successful fabrication
technologies~\cite{Russell2003}. PCFs are characterized by a
conventional cylindric geometry with a two-dimensional lattice of
air holes running parallel to the fiber axis. Such structures
share many properties of photonic crystals, associated with the
existence of the frequency gaps where the light transmission is
suppressed due to Bragg scattering, as well as the guiding
properties of conventional {\em optical fibers}, due to the
presence of a core in the structure.

Recent theoretical and experimental results reported the studies
and fabrication of dual-core PCF structures for broadband
directional coupling or polarization
splitting~\cite{p1,p2,p3,p4,p5}. In PCFs, light confinement is
restricted to the core of the fiber and therefore nonlinear
effects, such as light self-trapping and localization in the form
of spatial optical solitons~\cite{book}, become important. In
particular, similar to two-dimensional nonlinear photonic
crystals~\cite{Mingaleev2001}, a PCF can support and stabilize
both fundamental and vortex spatial optical
solitons~\cite{Ferrando2003,Ferrando2004}. In a sharp contrast
with an entirely homogeneous nonlinear Kerr medium where spatial
solitons are unstable and may collapse, it was shown that the
periodic structure of PCF can stabilize the otherwise unstable
two-dimensional solitons.

In this Letter, we make a further step forward in the study of
nonlinear effects in the PCF geometry, and analyze the existence
and stability of nonlinear guided modes and spatial solitons in
dual-core photonic crystal fiber couplers. The beam propagation
and power-dependent switching in nonlinear directional couplers
have been analyzed for the planar waveguide geometry~\cite{steg1}.
Here we generalize those results for PCFs, as well as study the
existence and stability of guided modes and {\em elliptic spatial
solitons} controlled by the PCF holes. In particular, we find
numerically the profiles of symmetric, antisymmetric, and
asymmetric nonlinear modes, and analyze all-optical switching
based on the mode instability.
\begin{figure}[htbp]
  \centerline{
    \scalebox{0.29}{
      \includegraphics{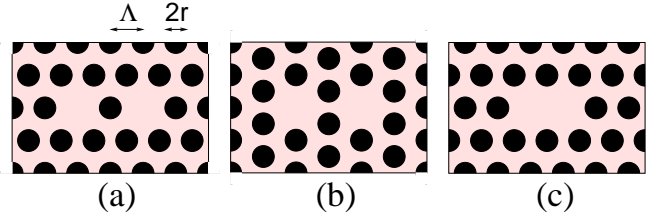}}
  }
\caption{(a-c) Schematic of three designs of a nonlinear dual-core
photonic crystal fiber. The coupler is created by two missing
neighboring holes in a nonlinear material.
  }
  \label{fig0}
\end{figure}

We consider a simple model of PCF that describes, at a given
frequency, the spatial distribution of light in a nonlinear
dielectric material with a triangular lattice of air holes with
the radius $r$ in a circular geometry. We assume that the PCF
material possesses a nonlinear Kerr response, and there are two
holes at the center filled by the same material creating a
nonlinear defect, as shown in Figs.~\ref{fig0}(a-c). In the
substrate material, the linear refractive index is $n_s$, whereas
inside the holes it is $n_a$. In the nonlinear regime, the light
distribution in PCF is described by the equation,
\begin{equation}
-i \frac{\partial E}{\partial z} = \Delta_{\perp} E+
     W(x,y) E + V(x,y)|E|^2E,
     \label{eq1}
\end{equation}
where $ W(x,y)= n_a+(n_s-n_a)V(x,y)$,
$\Delta_{\perp}=\partial^2/\partial x^2 + \partial^2/\partial y^2$
is the transverse Laplacian, $E$ is the normalized electric field,
and $V(x,y)$ is an effective potential describing two solid cores
in the lattice of holes. We normalize $V = 1$ in the material, and
$V = 0$ in the holes.

To find stationary nonlinear modes of PCF, we look for solutions
in the form $E(x,y,z)=u(x,y)\exp (i\beta z)$, and obtain the
nonlinear eigenvalue problem,
\begin{equation}
     \beta u = \Delta_{\perp} u+ W(x,y) u + V(x,y)|u|^2u.
     \label{eq2}
\end{equation}
In order to find the solutions of Eq.~(\ref{eq2}) for nonlinear
localized modes, we consider a rectangular domain of the $(x,y)$
plane and apply a finite-difference scheme, taking respectively
$N$ and $M$ uniformly distributed samples, $x_i,~ 0\leq i<N$ and
$y_j,~ 0\leq j<M$ of the variables, as well as the corresponding
samples for the stationary state, $u_{ij}=u(x_i,x_j)$ and the
potential $V_{ij}=V(x_i,y_j)$. Substituting these variables into
the model~(\ref{eq2}), and imposing homogeneous boundary
conditions in all four edges of the domain, we obtain an algebraic
nonlinear problem of $2\times N\times M$ equations with the same
number of unknowns $u_{ij}$, which is finally solved by means of a
globally convergent Newton method. The presence of the external
linear potential given by two missing holes and the lattice of
air-holes makes the system non-scalable and its radial symmetry
broken. Another approach, that takes an advantage of the lattice
periodicity, was developed recently by Ferrando {\em et
al.}~\cite{Ferrando2003}.

First, we consider a single missing hole and find numerically
solutions for the PCF spatial solitons~\cite{Ferrando2003}.
Importantly, these stationary solutions are not perfectly radial,
but they are stabilized by the PCF holes, in a sharp contrast with
the unstable self-trapped beams in nonlinear focusing Kerr media.
In order to demonstrate this feature, we follow the standard
analysis of the soliton stability~\cite{book} and analyze the
soliton power as a function of the soliton propagation constant. A
positive slope of this dependence indicates the soliton stability.
\begin{figure}[htbp]
  \centerline{
    \scalebox{0.75}{
      \includegraphics{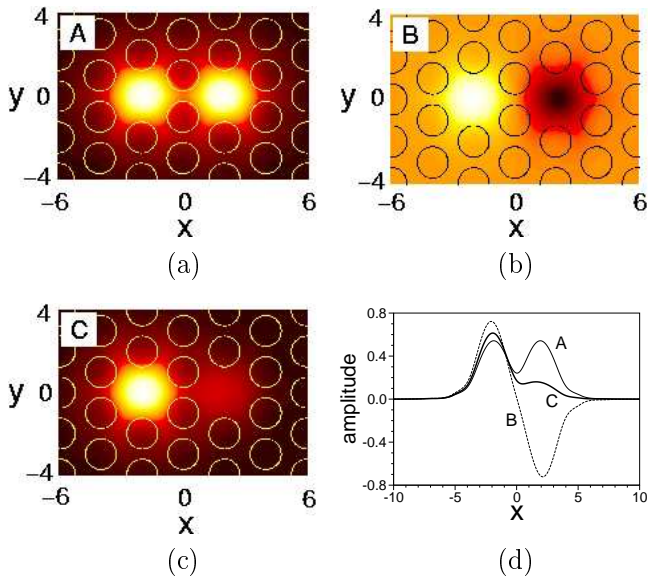}
    }
  }
  \caption{(a-c) Light distribution between the two
  PCF cores in Fig.~\ref{fig0}(b) for three distinct nonlinear modes:
    symmetric (A), antisymmetric (B), and asymmetric (C). (d) Transverse
    profiles of the nonlinear modes ($\beta=3.95$).}
  \label{fig1}
\end{figure}

Next, we study the dual-core nonlinear PCFs shown in
Figs.~\ref{fig0}(a,b), and find the families of the spatially
localized modes--the so-called PCF spatial solitons--as a function
of the mode propagation number $\beta$. The corresponding
solutions are similar for two cases of Figs~\ref{fig0}(a,b), and
they can be envisaged as the modes of the effective dual-core
fiber generated by the combined effect of the dual-core PCF
refractive index and the nonlinear index induced by the mode
amplitude itself. We find that the nonlinear dual-core PCF
supports three distinct nonlinear modes, symmetric mode (A),
antisymmetric mode (B), and asymmetric mode (C), as shown in
Fig.~\ref{fig1} for the case of Fig.~\ref{fig0}(b). The
corresponding spatial profiles of these modes are shown in
Fig.~\ref{fig1}(d) as cross-section cuts along the line $y=0$.
All the calculations are for $n_s=5$, $n_a=0$, $r=0.75$ and
$\Lambda=2$.
\begin{figure}[htbp]
  \centerline{
    \scalebox{0.35}{
      \includegraphics{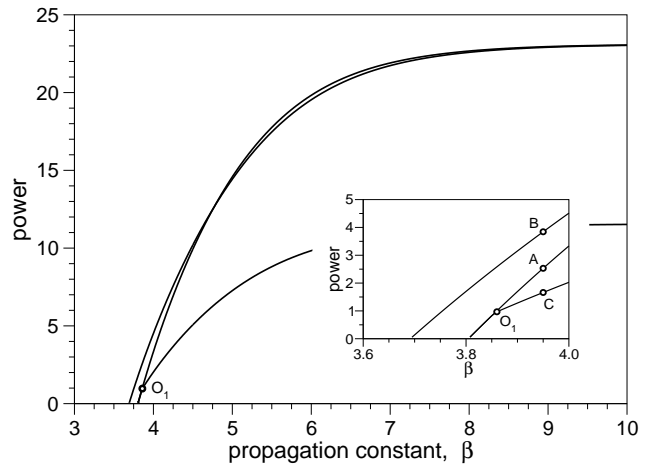}
    }
  }
  \caption{Bifurcation diagram of the coupler modes. Insert shows an enlarged part
  near the bifurcation point. Asymmetric mode C bifurcates from
  symmetric mode A above a certain threshold in the mode
  power.
  }
  \label{fig2}
\end{figure}

To demonstrate the power relation between the modes, we follow the
standard analysis of the soliton stability~\cite{book2} and plot
in Fig.~\ref{fig2} the soliton power as a function of the soliton
propagation constant, for the three different families. We notice
that only two modes, symmetric and antisymmetric ones, may exist
for low powers, whereas the asymmetric mode bifurcates from the
symmetric mode at a ceratin threshold value of the mode power,
above which the symmetric mode becomes unstable.
\begin{figure}[htbp]
  \centerline{
    \scalebox{0.32}{
      \includegraphics{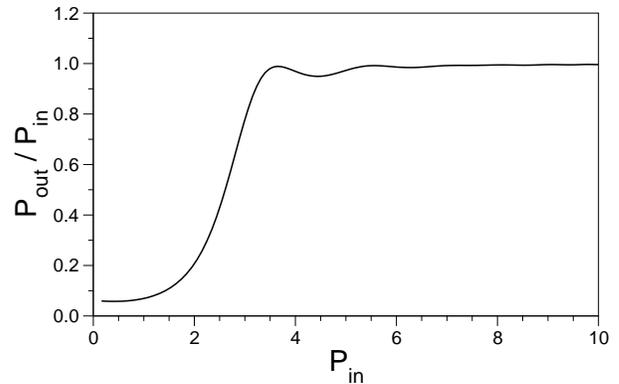}
    }
  }
  \caption{
Switching curve calculated for the PCF nonlinear coupler of
Fig.~\ref{fig0}(b). Due to instability of the symmetric mode, the
light launched into one core only does not switch to the second
core but remains in the same core.
  }
  \label{fig3}
\end{figure}

\begin{figure}[htbp]
  \centerline{
    \scalebox{0.7}{
      \includegraphics{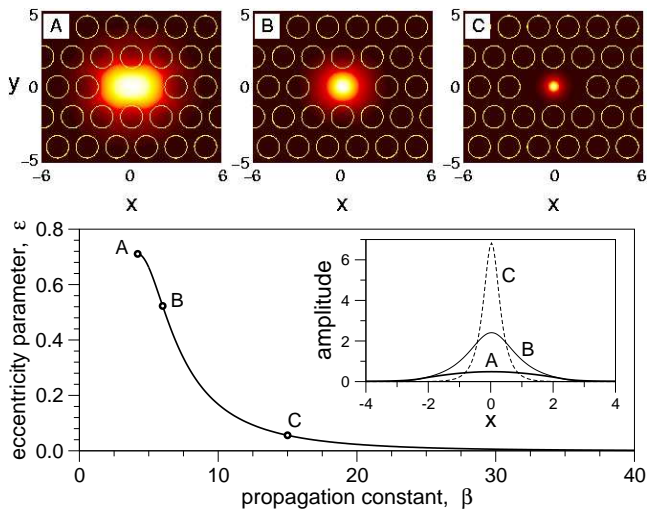}
    }
  }
  \caption{Transformation of the elliptic guided modes to circular
solitons for the growing mode power. Three examples on the top
mark three points of the eccentricity curve for $\beta=4.2$,
$\beta=6$, and $\beta=15$ respectively. Inset: the corresponding
beam cross-sections at $y=0$.
  }
  \label{fig4}
\end{figure}


In order to study the switching properties of the nonlinear
coupler, we carry out a series of numerical simulations using the
standard beam propagation method. First, we calculate the
stationary mode of a single-core PCF with the same parameters as
the coupler corresponding to different powers. Then, this mode is
launched into one of the cores of the dual-core coupler and, after
some propagation, the output power is calculated at the same core.
The propagation distance is determined considering the same
coupler operating in the linear regime and propagating the linear
single-core mode launched to one of the coupler cores up to the
point where all the energy is transferred to the second core. For
the parameters used in our examples this point is reached at
$z=21.23$. The switching curve is obtained by varying the input
power for the fixed propagation length, and it is plotted in
Fig.~\ref{fig3}. Similar to other types of nonlinear directional
couplers, the input power is transferred completely to the second
core for low powers, while for it remains in the initial one for
higher input powers. A change between both the regimes takes place
in a relatively short range of powers which constitutes a
threshold where the power switching is triggered. For very low
powers the energy is completely transferred to the second core,
although from the plot in Fig.~\ref{fig3} a residual amount seems
to remain in the first one. This nonzero behavior of the curve
close to the origin is explained by the overlapping of the field
of the second core due to its proximity.

Finally, we study the third case of the closely spaced holes,
shown in Fig.~\ref{fig0}(c). In this case, there exists no
bifurcation to the asymmetric state and the fundamental mode
itself is elliptic as shown in the inset of Fig.~\ref{fig4}. In
the nonlinear case, this elliptic guided mode gives a birth to an
elliptic spatial soliton controlled by the boundaries of the
holes.  In Fig.~\ref{fig4}, we plot the mode eccentricity
parameter $\epsilon=[1-(w_x/w_y)^2]^{1/2}$, where $w_x$ and $w_y$
are the mode axes (widths), and thus quantify the transformation
of the elliptic guided modes (A) to the elliptic solitons (B) and
then to the radially symmetric solitons (C). These results
resemble the transformation of the shape of nonlinear guided modes
in planar waveguides~\cite{sammut}.

In conclusion, we have demonstrated that several types of
two-dimensional spatial optical solitons can be supported by a
nonlinear dual-core PCF structure with the Kerr nonlinearity. We
have analyzed numerically the existence and stability of
symmetric, antisymmetric, and asymmetric nonlinear modes
demonstrating that periodic refractive index of PCF provides also
an effective stabilization mechanism for these composite localized
modes to exist in a nonlinear Kerr medium, in a sharp contrast
with an entirely homogeneous nonlinear Kerr medium where spatial
solitons are known to be unstable, and they undergo the collapse
instability. We have studied all-optical switching in the
nonlinear dual-core PCF coupler associated with the instability of
the symmetric mode.

The authors thank Adrian Ankiewicz for useful discussions and
acknowledge a partial support of the Australian Research Council.
JRS acknowledges a visiting fellowship granted by the Direcci\'on
Xeral de Investigaci\'on e Desenvolvemento of Xunta de Galicia
(Spain), and he thanks Nonlinear Physics Center for a warm
hospitality during his stay in Canberra.

$^{\dagger}$On leave from: Facultade de Ciencias de Ourense,
Universidade de Vigo, 32004 Ourense, Spain.


\begin{thebibliography}{}

\bibitem{Russell2003} Ph. Russell, Science {\bf 299}, 358 (2003)

\bibitem{p1} F. Fogli, L. Saccomandi, P. Bassi, G. Bellanca, and
S. Trillo, Opt. Express {\bf 10}, 54 (2002).

\bibitem{p2} L. Zhang and Ch. Yang, Opt. Express {\bf 11}, 1015
(2003).

\bibitem{p3} K. Saitoh, Y. Sato, and M. Koshiba, Opt. Express {\bf
11}, 3188 (2003).

\bibitem{p4} H. Kim, J. Kim, U.-C. Paek, B.H. Lee, and K.T. Kim, Opt. Lett. {\bf 29}, 1194 (2004).

\bibitem{p5} J. Lagsgaard, O. Bang, and A. Bjarklev, Opt. Lett.
{\bf 29}, 2473 (2004).

\bibitem{book} Yu.S. Kivshar and G.P. Agrawal, {\em Optical Solitons:
From Fibers to Photonic Crystals} (Academic, San Diego, 2003), 540
pp.

\bibitem{Mingaleev2001} S.F. Mingaleev, and Yu.S. Kivshar, Phys. Rev. Lett.
{\bf 86}, 5474 (2001).

\bibitem{Ferrando2003} A. Ferrando, M. Zacar\'es, P. Fernandez de C\'ordoba, D. Binosi,
 and J.A. Monsoriu, Opt. Exp. {\bf 11}, 452 (2003).

\bibitem{Ferrando2004} A. Ferrando, M. Zacar\'es, P. Fernandez de C\'ordoba, D. Binosi,
 and J.A. Monsoriu, Opt. Exp. {\bf 12}, 817 (2004).

\bibitem{steg1} L. Thylen, E.M. Wright, G.I. Stegeman, C.T.
Seaton, and J.V. Moloney, Opt. Lett. {\bf 11}, 739 (1986).

\bibitem{book2} N.N. Akhmediev and A. Ankiewicz, {\em Solitons} (Chapman \& Hall, London, 1997).

\bibitem{sammut} Q.Y. Li, C. Pask, and R.A. Sammut, Opt. Lett. {\bf 16}, 1083 (1991).

\end{thebibliography}
\end{document}